# How Astronomers Perceive the Societal Impact of Research: An Exploratory Study

Michelle Willebrands
Leiden University
willebrands@strw.leidenuniv.nl

Pedro Russo
Leiden University
russo@strw.leidenuniv.nl

## Summary

We present an exploratory study of the perception of professional astronomers about the societal impact of astronomy. Ten semi-structured interviews with astronomers from a range of career and cultural backgrounds have been conducted to gain in-depth insight into their opinion about societal impact and their approach in realising it. The results show that the interviewees are aware of the diversity of impacts that astronomical research has. However, they are mostly active in outreach and only a few activities are incorporated into their jobs to achieve an impact on development. There is little contact with stakeholders in industry, policy or other fields, like development. Besides, a structured approach in their personal outreach is lacking, and assessment is only done informally. Despite the limited sample size of this study, the results indicate that a further change is necessary to engage professional astronomers with topics of development and societal impact to create action on the level of individual researchers.

Body Text:

## Introduction

In most research institutes, there are three main aims: research, teaching and public engagement. However, the latter is often neglected, although in the past years there has been a growing emphasis on the societal impact of science. The societal impact of science can be understood as science that includes societal benefit and affects societal challenges (Bornmann, 2013). It entails social, cultural, environmental and economic returns and engages societal actors, such as policymakers, industry and end users.

Astronomy impacts societies in different ways by producing certified knowledge; training skilled workers; driving innovation by pushing technical limits; contributing to collective goods, like prestige of a country and environmental awareness; and by inspiring people (Davoust, 1995). Serendipity is extremely important in astronomical discoveries (Fabian, 2010), and societal applications of science are often arrived at non-linearly (Schneider, 2007). More spin-offs might even come from fundamental than applied research (Llewellyn Smith, 2008). Rosenberg et al. gathered a wealth of examples of applications originating in astronomy, from X-ray luggage belts to hospital cleanrooms. However, they conclude that maybe the most important consequence of astronomy is that it highlights our place in the universe and promotes global citizenship. Astronomy can also contribute to socioeconomic development (McBride, 2018).

**Background**

The process of creating societal impact can be captured in a four-step process (Meijer, 2012). Firstly, the societal objectives are defined, and subsequently the stakeholders and activities to connect with them. Next, the impact must be measured with indicators, and finally, the results are reflected by scoring each indicator to adjust the objectives if necessary.

Outreach efforts of physicists and biologists mostly concern presentations for children and activities for a general audience, like public lectures (Ecklund et. al., 2012). Perceived impediments to outreach activities include the "Sagan effect" (Hartz et. al., 1997), where individuals who do outreach are thought to do less rigorous research by peers. Besides, researchers believe the public is disinterested in science and there is doubt whether scientists or an intermediary should be responsible for outreach. Finally, institutes prioritise research and there is often little time to engage in outreach as well as a lack of reward for it (Ecklund).

Most astronomers have a positive attitude towards education and outreach, although they spend less time on it than recommended (Dang, 2015). Personal motivations are the main drive to interact with the public but there is little institutional support (Sarperi et. al., 2018), with seniority being another important factor (Entradas et. al., 2018).

### Research Question

The main goal of this article was to explore what professional astronomers perceive to be the societal impact of astronomy and whether they incorporate it into their work. Previous research has been conducted on how astronomers engage with the public, as well as their motivations for it (see Ecklund; Dang; Sarperi; Entradas; and Bastow, 2014). However, there is little research on astronomers' concern about the wider impact of their work on society and if there is a systematic approach.

We presume that there is a discrepancy in the attitude of astronomers: they might believe that societal impact is important, but not incorporate it into their work. We expected the main method used to affect society are traditional outreach activities without defining a broader strategy.

Many factors might play a role in how any astronomer regards societal impact (e.g. seniority, field of expertise, socio-geographical-cultural demographics, personality traits). To mitigate the influence of these factors, the interviewees were carefully selected. However, the effect of such factors could not be investigated quantitatively. This is a small-scale exploratory study that should be built on in future research.

**Data Collection and Analysis**

The interviewees were selected to provide the best representation possible of the global professional astronomy community within the limited sample size. Four face-to-face interviews took place during the General Assembly of the International Astronomical Union in 2018 and six were conducted online. Details about the sample of interviewees is provided in Table 1.

**Table 1.** Characteristics of the participant sample.

| Gender | Age | Nationality | Based in | Academic position | Field |
|---|---|---|---|---|---|
| F | 67 | UK | South Africa | Professor | Variable stars |
| F | 40 | Lebanon | Lebanon | Assistant professor | Mercury exosphere, comets |
| M | 36 | Russia | USA | Assistant professor | Computational/theoretical astrophysicist |
| M | 30 | India | USA | Postdoc | Extragalactic, galaxy dynamics |
| M | 49 | USA | The Netherlands | Professor | Galaxy evolution, stellar populations of galaxies, instrumentation |
| M | 59 | Japan | Japan | Professor | Radio astronomy, Milky Way |
| M | 43 | UK | UK | (PhD) Director of corporate strategy | (Pola Aurora) Policy and strategy |
| M | 61 | Thailand | UK | (Postdoc) Diplomat science and technology | (Geology, crystallography) Science policy and communication |
| M | 41 | Chile | Chile | Associate professor | Planetary atmospheres |
| F | 24 | Australia | The Netherlands | PhD student | Galaxy formation and evolution |

Ten semi-structured interviews were carried out to allow for comparison between interviewees and identify common narratives, and to explore relevant topics outside the interview questions (DiCicco-Bloom et. al., 2006). "Professional astronomer" in this study is defined as a person who is affiliated with an astronomy research institute.

The interviews were recorded after informed consent from the participant. A list of questions formed the framework of the interview and care was taken to formulate questions in an open and non-leading manner. The interviews were structured around the interview protocol:

- Background of the participant
- Attitude towards societal impact of astronomy
- Attitude of their institute and colleagues
- Knowledge about different types of impact
- Approach in achieving societal impact
- Barriers in this approach and possible improvements

The recordings of the interviews were transcribed and common topics in the data were grouped in codes. These codes were assigned to overarching categories (Figure 1). After the codes were determined, the transcripts were analysed again and re-labelled if necessary to ensure rigorous data analysis.

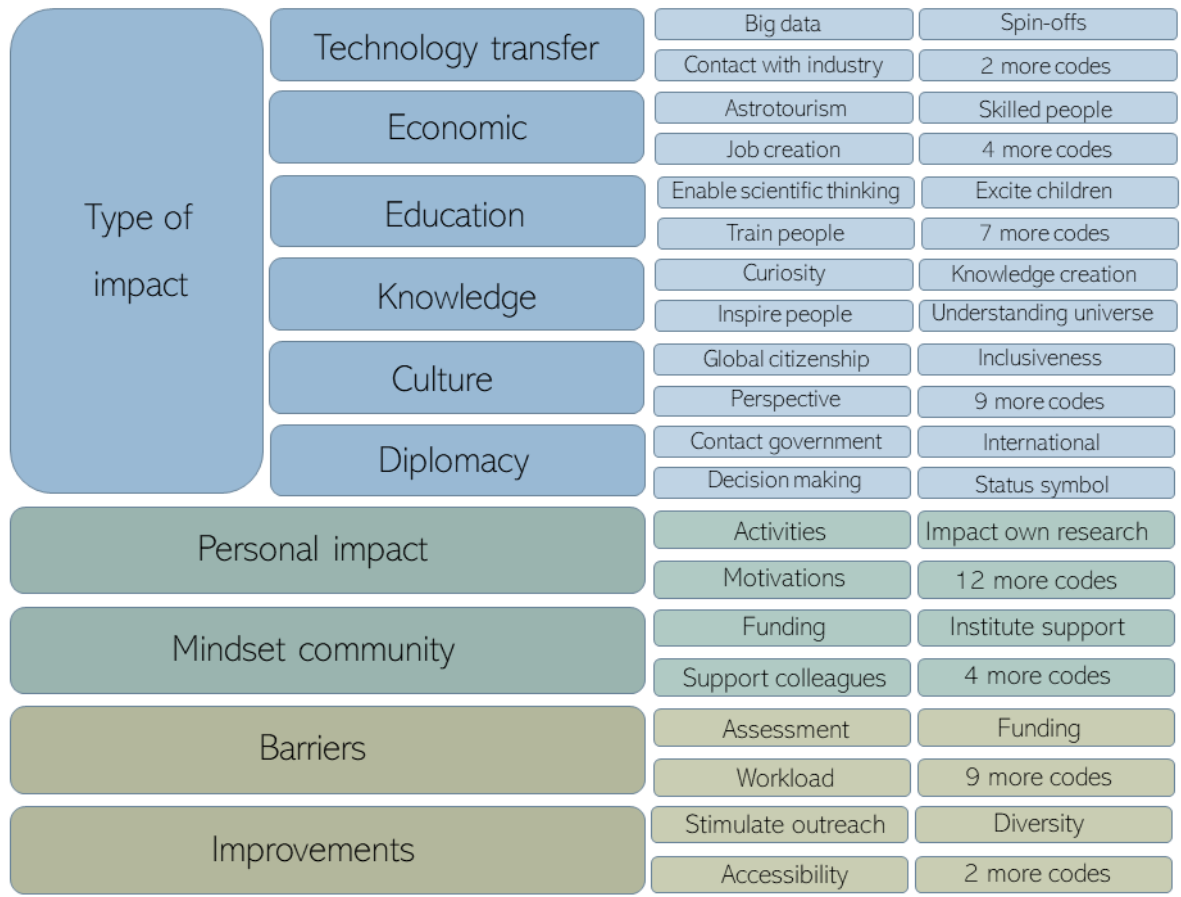

**Figure 1.** A visualisation of the categories and codes obtained in the data analysis of the interview transcripts. The categories (left column) consist of six types of impact, personal impact, the mindset of the community, barriers and opportunities for improvements reported by the interviewees. A selection of the corresponding codes (middle and right column) are provided for each category. Image credit: M. Willebrands

## Results

### Types of Impact

Based on the examples mentioned by interviewees, the authors categorised six types of societal impact of astronomy: technology transfer, economic returns, education, knowledge creation, cultural dimension, and diplomacy (Table 2).

**Table 2.** Overview of the types of impacts of astronomy on society mentioned by the interviewees, per category. The right column shows the number of interviewees out of the sample of 10 that mentioned each impact.

| Category | Impact | N = x |
|---|---|---|
| Technology transfer | Spin-offs<br>Big data | 10<br>5 |
| Economic benefits | Acquired skills<br>Astrotourism<br>Job creation | 7<br>5<br>2 |
| Education | Attract children to science<br>Critical thinking<br>Counteract misinformation<br>Human capital | 8<br>7<br>6<br>5 |
| Knowledge creation | Inspire<br>Understanding the world<br>Societal advancement | 10<br>7<br>2 |
| Cultural | Perspective<br>Inclusiveness and diversity<br>Global citizenship<br>Appreciation of the Earth | 6<br>5<br>3<br>3 |
| Diplomacy | International collaboration<br>Decision-making<br>Research infrastructures | 4<br>3<br>2 |

### Societal Impact Efforts

All the interviewees engage in outreach activities that they feel contribute to society and some (N=5) feel that it is their duty to do so. Besides sharing discoveries in media, many participants give public talks. However, they agree that their own research generally does not have a real impact on society (N=9).

Despite the active role taken by the participants, some (N=3) question who is responsible, researchers or an intermediary, for achieving impact. Furthermore, one interviewee thinks societal impact should not always be the main goal of research and that "there has to be room for things that are just interesting, impact does not need to drive it".

All participants agree that societal impact is emphasised nowadays and it cannot be ignored. Seven of them know that their institutes include it in their mission. However, the answers to whether public engagement activities are appreciated are conflicted. The participants indicate that engaging in activities related to societal impact is generally appreciated by colleagues, although there is not always respect for it.

The sample expresses mixed opinions on whether the professional astronomy community wants to actively pursue societal impact. One participant believes that many astronomers like to do outreach, while another thinks that most are "worried about bugs in their code, rather than talking to the public". If given the opportunity to do public engagement most would do it, but they would not seek it out.

### Barriers and Improvements

Generally, the participants realise that creating societal impact with fundamental science is an unpredictable process (N=6) as *"it's hard to know what the impact will be of what you're studying"*. They mostly feel like they can spend time on societally-relevant activities within their job (N=6). Participants indicate that a lack of funding is an obstacle to achieving societal impact (N=4), as well as the high workload of their job (N=3) and language (N=2). One interviewee mentions the competition between universities as an obstruction.

The main reported improvement is the inclusion of minorities and diversity (N=5). Some participants (N=4) mention the accessibility of astronomy ("Science needs to feel familiar to people."), and only one participant has concerns about potential negative impact and emphasises the need for two-way communication ("Try to have a dialogue with them, not just explain."). Assessment of societally-impactful activities that the participants engage in is often only done informally and without structure (N=7).

**Conclusion**

The professional astronomers in this study are aware of the different ways astronomy can impact positively on society. They deem societal impact important and are motivated to communicate with the general public, mainly through talks and classroom activities, and feel like they have time for doing so. Beyond the general public, they have few connections with industry or policymakers. Self-reported barriers in achieving societal impact include high workload, priority on research tasks, lack of funds and language barriers. Besides, they realise that societal impact is difficult to measure due to the serendipitous nature of discoveries in astronomy (Ecklund, 2012). According to the participants, astronomers could improve on achieving impact by making research more accessible and being more inclusive with underrepresented groups.

The activities that the interviewees undertake to achieve impact are mostly one-way exchanges with the general public and there are little efforts to interact with non-scientific stakeholders. Moreover, they encounter barriers and there is no structured approach in their public engagement activities like for research. The participants do not generally think about the impact they want to have and do not formally assess the outcomes. They know about the potential societal impacts of astronomy but do not have all the tools to incorporate it into their work.

Even though no rigorous conclusions can be drawn based on the limited sample size, the results indicate that professional astronomers might not incorporate adequate activities into their job to achieve societal impact. Training astronomers to adopt a backwards approach, where the desired societal impacts of an activity are defined first, could be part of the solution.

**Biographies**

Michelle Willebrands is the project manager of the International Astronomical Union's (IAU) European Regional Office of Astronomy for Development (E-ROAD) at Leiden Observatory. She is the lead on the Advocacy and Legacy work package of the H2020 spaceEU project.

Pedro Russo is university professor in Astronomy & Society at Leiden University, the Netherlands. He coordinates the Astronomy & Society group at Leiden Observatory, which implements global-scale projects in astronomy, space education and public engagement.